\documentclass{elsart}

\usepackage{graphicx}

\begin{document}
\begin{frontmatter}
\title {LED Monitoring System for the BTeV Lead Tungstate Crystal 
Calorimeter Prototype}
\author[IHEP]{V.A.~Batarin},
\author[FNAL]{J.~Butler},
\author[NAN]{T.Y.~Chen},
\author[IHEP]{A.M.~Davidenko},
\author[IHEP]{A.A.~Derevschikov},
\author[IHEP]{Y.M.~Goncharenko},
\author[IHEP]{V.N.~Grishin},
\author[IHEP]{V.A.~Kachanov},
\author[IHEP]{A.S.~Konstantinov},
\author[IHEP]{V.I.~Kravtsov},
\author[IHEP]{V.A.~Kormilitsin},
\author[UMN]{Y.~Kubota},
\author[IHEP]{Y.A.~Matulenko},
\author[IHEP]{V.A.~Medvedev},
\author[IHEP]{Y.M.~Melnick},
\author[IHEP]{A.P.~Meschanin},
\author[IHEP]{N.E.~Mikhalin},
\author[IHEP]{N.G.~Minaev},
\author[IHEP]{V.V.~Mochalov},
\author[IHEP]{D.A.~Morozov},
\author[IHEP]{L.V.~Nogach},
\author[IHEP]{A.V.~Ryazantsev\thanksref{addr}},
\thanks[addr]{corresponding author, email: ryazantsev@mx.ihep.su}
\author[IHEP]{P.A.~Semenov},
\author[IHEP]{V.K.~Semenov},
\author[IHEP]{K.E.~Shestermanov},
\author[IHEP]{L.F.~Soloviev},
\author[SYR]{S.~Stone},
\author[IHEP]{A.V.~Uzunian},
\author[IHEP]{A.N.~Vasiliev},
\author[IHEP]{A.E.~Yakutin},
\author[FNAL]{J.~Yarba}
\collab{BTeV electromagnetic calorimeter group}
\date{\today}

\address[IHEP]{Institute for High Energy Physics, Protvino, Russia}
\address[FNAL]{Fermilab, Batavia, IL 60510, U.S.A.}
\address[NAN]{Nanjing University, Nanjing, China}
\address[UMN]{University of Minnesota, Minneapolis, MN 55455, U.S.A.}
\address[SYR]{Syracuse University, Syracuse, NY 13244-1130, U.S.A.}

\begin{abstract}
     We report on the performance of a monitoring system for a prototype 
calorimeter for the BTeV experiment that uses lead tungstate crystals 
coupled with photomultiplier tubes. The tests were carried out at the 
70 GeV accelerator complex at Protvino, Russia.
\end{abstract}

\begin{keyword}
light emitting diode \sep monitoring system \sep stability \sep calorimeter
\sep scintillating crystal
\PACS 29.90.+r \sep 85.60.Dw \sep 85.60.Jb \sep 85.60.Ha
\end{keyword}

\end{frontmatter}

\section{Introduction}

The BTeV experiment at Fermilab will use
an electromagnetic  calorimeter (EMCAL) made of lead tungstate 
(PbWO$_4$) crystals \cite{prop}.
These are scintillating crystals with a rather complex emission
spectrum, consisting of two emission components: blue, peaking at 420 nm 
and green, peaking at 480-520 nm. Most of the light, $99\%$, is emitted in 100 ns.
The properties of the crystals produced by different
manufacturers as well as the characteristics of the EMCAL prototype made of these
crystals have been investigated using a test-beam facility at the 70-GeV accelerator
at Protvino, Russia. Results of the measurements
as well as a detailed description of the test-beam facility are given elsewhere
\cite{nim1}, \cite{nim2}, \cite{nim3}.  

The light output of the PbWO$_4$ crystals is reduced, as a rule, when they are 
irradiated using electron and pion beams.
The main reason for this effect is thought to be a lowering of the crystals 
transparency in a wavelength dependent manner. To study the magnitude of this
effect we constructed a light output monitoring system that used
light emitting diodes (LED) at four different wavelengths covering the range 
from 400 to 660 nm. 
The most important characteristics of this system include: 
easy adjustment of the light pulse duration and intensity, high reliability, 
low power consumption, durability, and finally low cost.

The BTeV lead tungstate crystals will be continuously calibrated 
{\em in situ} at the Tevatron using different physical processes.
It is foreseen that electrons from $B$ decays as well as photon conversions
will be used for this purpose.
The overall system  must be able to track the light output changes in each 
crystal to an accuracy of $0.2\%$. 
The time required to collect enough events for the energy calibration varies
from less than an hour for the most hit $10\%$ of the crystals
laying close to the beam to about 10-20 hours, for the least hit crystals.
A light monitoring system will track the transparency variation over these time
intervals in order to guarantee that we maintain the calorimeter's energy resolution.

The monitoring system described here has already proven to be an invaluable 
tool for our systematic study of the crystal properties at the test-beam 
facility.
The main goal of the present study is to measure the levels of instability
of this system. This information will allow us to  decide if this type of 
system can be used in the final design of the BTeV monitoring system. 
 
The stability of the monitoring system was evaluated using 
data collected from special LED pulse triggers intermixed with data taken 
using intense beams that served to irradiate the PbWO$_4$ crystals. 
The crystals were exposed to the beam in December 2002
and recovery was monitored between January and March of 2003.

\section{Test-beam Facility}

The BTeV calorimeter test-beam setup consisted of a 5$\times$5 array of lead
tungstate crystals coupled to photomultiplier tubes, a beam
with a momentum tagging on individual particles and a trigger 
system using scintillation counters. To eliminate the effects of
temperature variation, crystals were placed inside a thermally
insulated light-tight box. The temperature was measured continuously
at 24 different locations around the crystal array using thermo-sensors.
A more detailed description is given in \cite{nim2}.
The main difference from
our earlier test-beam studies is the use of 6-stage R5380Q Hamamatsu PMT's
(instead of 10-stage R5800 PMT's) for the crystals readout. This phototube is
one of the possible candidates to be used in the BTeV EMCAL. 
Signals from the PMT's were amplified by a factor of 10 using
electronics developed at Fermilab
to match the range of the LeCroy 2285 15-bit integrating ADC modules.
The amplifiers were placed near the PMT's inside the thermo-insulated box. 
The signal charge (either from particles or from LED's) was integrated 
over a 150 ns gate.

\section{Monitoring LED Pulser System Design}

The LED-based monitoring system was
designed to study variations of the crystals transparency 
while they were irradiated by high energy particles.
Because we used different wavelengths of light and the crystals transparency change 
under radiation is wavelength dependent, it also is possible to monitor changes of 
the PMT's gains.
 These can arise from changes in the average anode currents, 
as well as other reasons such as variations in the high voltage (HV) power supply.

The following LED's were used in the monitoring system:
\begin{itemize}
\item violet (Kingbright L2523UVC), peak wavelength at 400 nm;
\item blue (Nichia NSPB 500S), peak wavelength at 470 nm;
\item yellow (Kingbright L-53SYC), peak wavelength at 590 nm;
\item red (Kingbright L-53SRC-E), dominant wavelength at 640 nm.
\end{itemize}

Violet and blue LED's provided the main results about the crystal transparency
change in their respective wavelength regions.  
The transmission of red light in the PbWO$_4$ crystal is not 
affected much by radiation damage \cite{red}, and thus we use the red LED
to monitor the photomultiplier gains. This proved to be extremely valuable.

\begin{figure}
\centering
\includegraphics[width=0.95\textwidth]{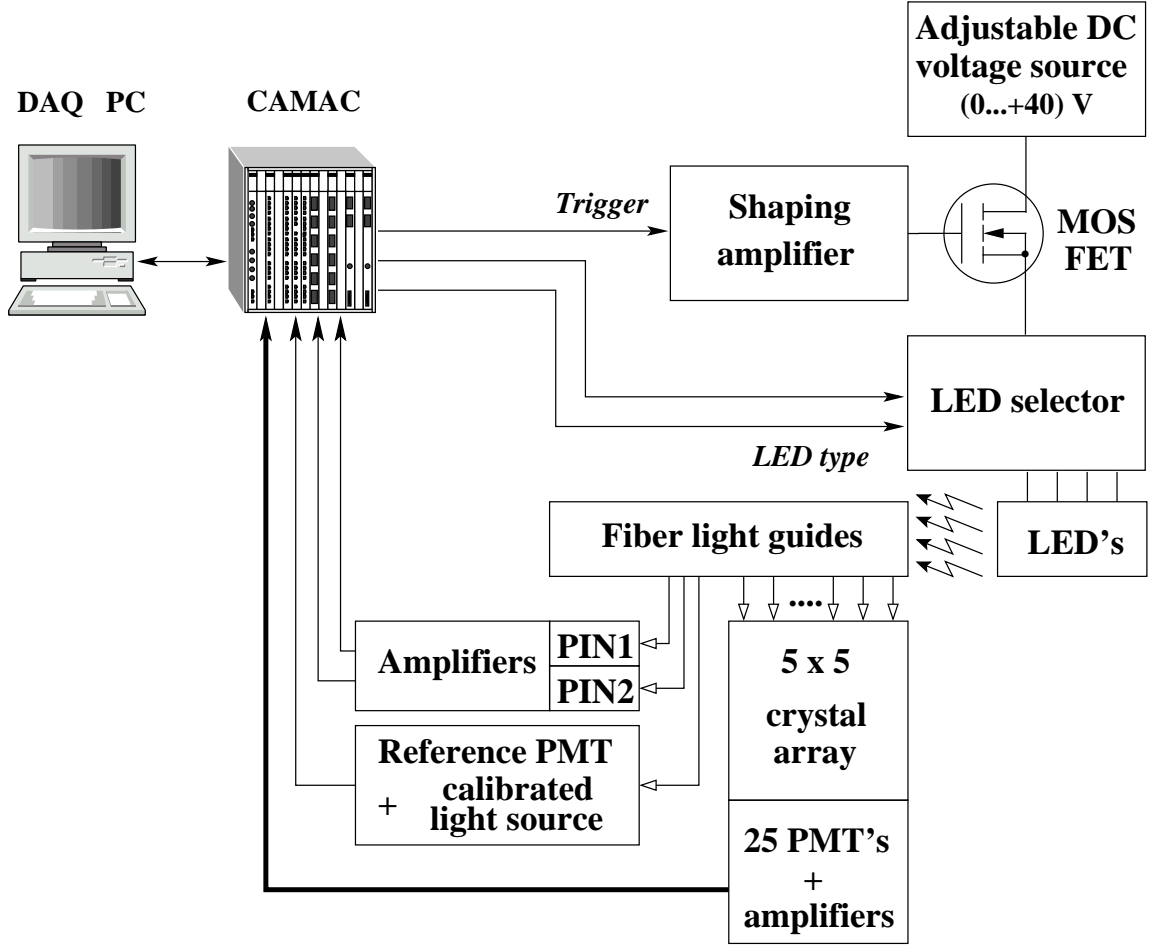}
\caption{Block diagram of the LED monitoring system.}
\label{fig:pulser}
\end{figure}

The system consisted of three main parts: a program-controlled LED
pulser, a distribution network comprised of optical fibers and a stability
monitoring subsystem. The block diagram is shown in Fig.~\ref{fig:pulser}.
All the components of the light monitoring system, except the adjustable direct
current (DC) voltage source, were placed inside the temperature stabilized box 
near the crystal array.

The LED pulser includes a  shaping amplifier and output transistor (MOS FET) 
in a switching mode. The FET source is connected to one of the four LED's using 
the LED selector, while the drain of the FET was connected to the stable 
DC voltage source.
The shaping amplifier determines the duration (100 ns) of the LED driving pulses.
The LEDs' capacitances made the duration of light pulses longer, 
but the signals from all four LED's were still shorter than the ADC gate width
of 150 ns.
The source voltage defines the light pulse intensity.
It can be set at any value between 0 and +40 V.

The LED pulser's operation mode was controlled remotely via the data 
acquisition system (DAQ) and can be easily modified if necessary.
The selected LED produced a series of 10 light pulses between 
two accelerator spills.  
After the next beam spill, an LED of another wavelength produced 10 pulses.
As a result, four accelerator cycles
were necessary to complete the readout of all four LED's.
The cycle duration is about 10 seconds,
therefore about 60 amplitude values per minute (15 for each LED) from each
photodetector were recorded during data taking. This provided enough statistics
for accurate monitoring every few minutes. 

\begin{figure}
\centering
\includegraphics[width=0.95\textwidth]{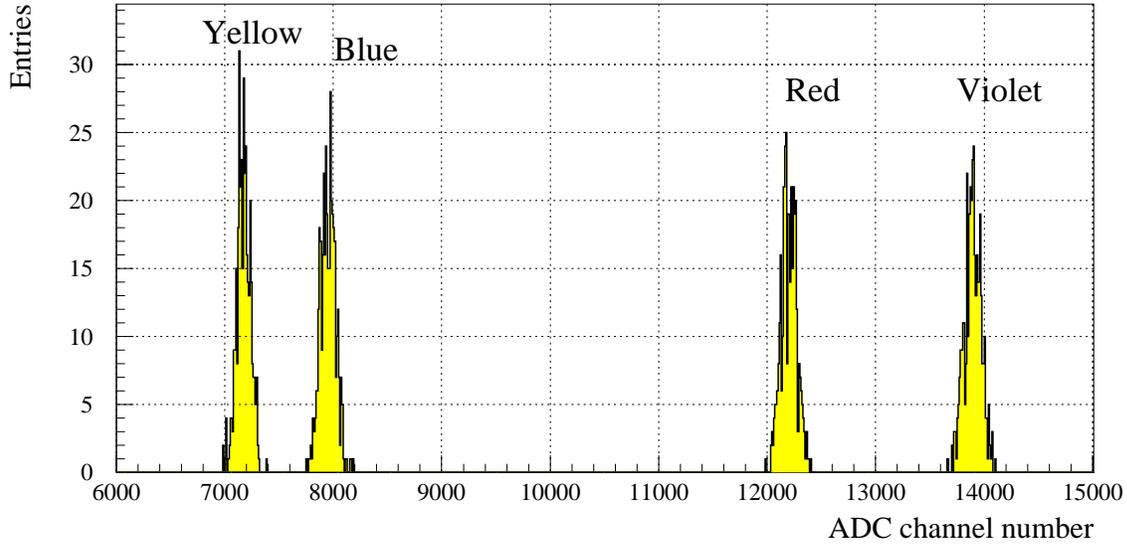}
\caption{Amplitude spectra in one of the calorimeter prototype channels from
four different LED's collected over 20 minutes. The r.m.s./peak ratios
are: $0.85\%$ for yellow,
$0.83\%$ for blue, $0.60\%$ for red and $0.53\%$ for violet.}
\label{fig:LED_cell}
\end{figure}

Clear plastic fiber light guides  were used to transmit light
from the LED's to the cells of the EMCAL prototype and to the stability monitoring
subsystem. LED's illuminated a bunch of optical fibers.
They were placed about 100 mm apart to provide a uniform illumination over 
the entire fiber bundle.
Fibers were attached to the far (from PMT's) ends of the crystals.
Typical pulse height distributions of the 
signals in one of the EMCAL prototype channels, produced
by four LED's of different wavelengths, collected over
20 minutes are shown in Fig.~\ref{fig:LED_cell}. 
The channel 10000 of ADC approximately corresponds to the peak position 
of the 20 GeV electron signal amplitude distribution.
 
\begin{figure}[b]
\begin{tabular}{lr}
\centering
\includegraphics[width=0.45\textwidth]{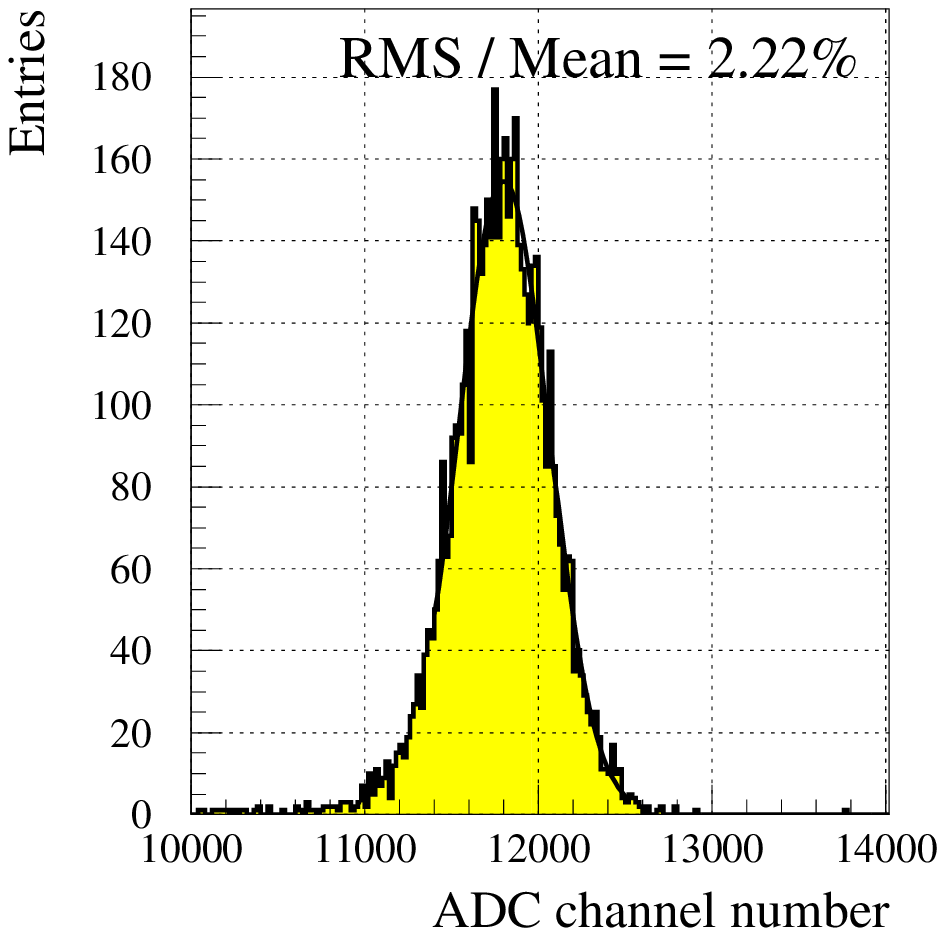}
&
\includegraphics[width=0.45\textwidth]{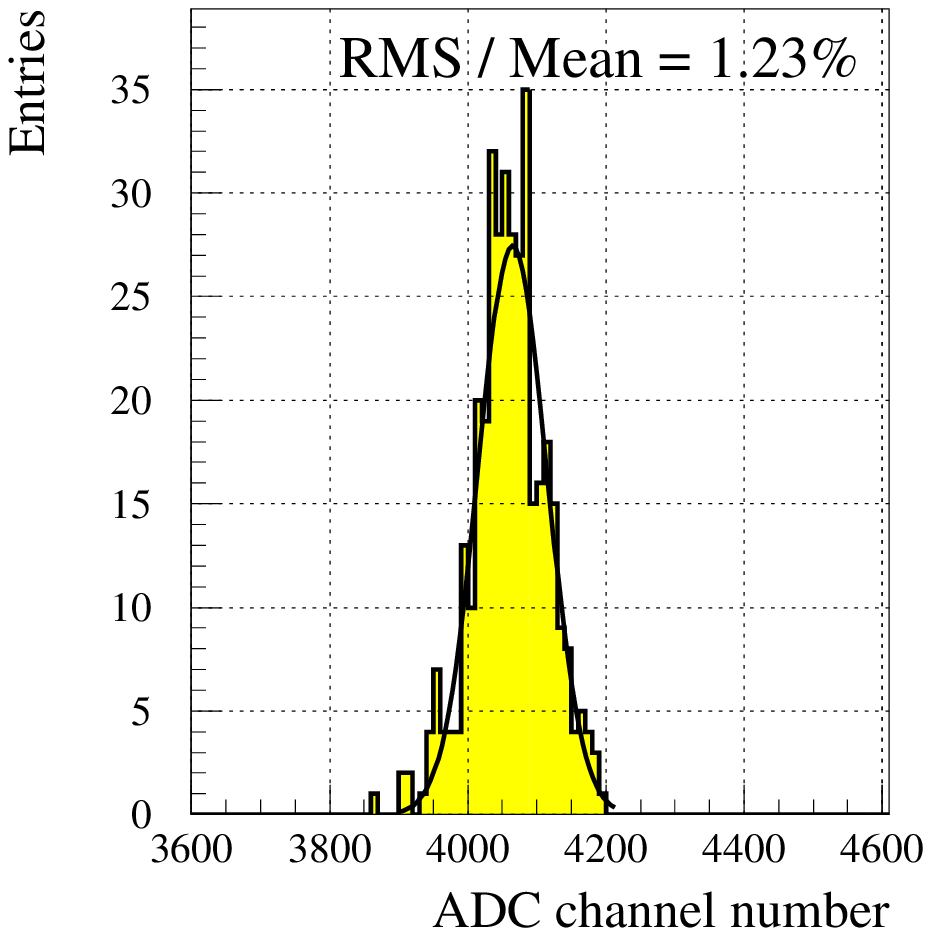}\\
\parbox{0.45\textwidth}{\caption{Amplitude spectrum from the reference 
light pulser (YAP:Ce crystal irradiated by $\alpha$-source) in the monitoring  PMT
collected over 20 minutes.\label{fig:alpha}}}&
\parbox{0.45\textwidth}{\caption{Amplitude distribution of blue LED signal in 
one of the PIN photodiodes collected over 20 minutes.\label{fig:LED_pin}}}\\
\end{tabular}
\end{figure}

To monitor the stability of the LED's we used two silicon PIN photodiodes Hamamatsu
S6468-05 and a PMT Hamamatsu R5800 as a reference photomultiplier tube with a calibrated 
light source mounted on its front window. This light source was comprised of a small
YAP:Ce crystal ($3\times3\times0.1$ mm$^3$) assembled in a plastic case
with an $\alpha$-source \cite{yap}. It provided about 20 flashes    
per second with maximum of emission spectrum at 360 nm and decay time of about 30 ns.
A signal from the PMT's last dynode was used for trigger. The DAQ recorded about 50
$\alpha$-events between two accelerator spills.  
An amplitude spectrum obtained from this reference light source is
shown in Fig.~\ref{fig:alpha}.

The S6468-05 is a photodiode and a preamplifier chip integrated in 
the same package. It has an active area of 0.8 mm diameter and good 
sensitivity over a wide spectral range from 320 to 1000 nm. 
Two photodiodes were mounted on a small printed circuit board with 
additional AD8002 integrated circuit--based amplifiers and 
two voltage stabilizer integrated circuits, supplied $\pm5 V$ DC voltage
for the amplifiers.
Figure~\ref{fig:LED_pin} shows a typical amplitude distribution of 
blue LED signals obtained from one of the PIN photodiodes. The width of 
this distribution is caused mainly by the noise of the amplifiers 
rather than photon statistics.

\section{Stability Analysis Method}

\begin{figure}[b]
\centering
\includegraphics[width=0.95\textwidth]{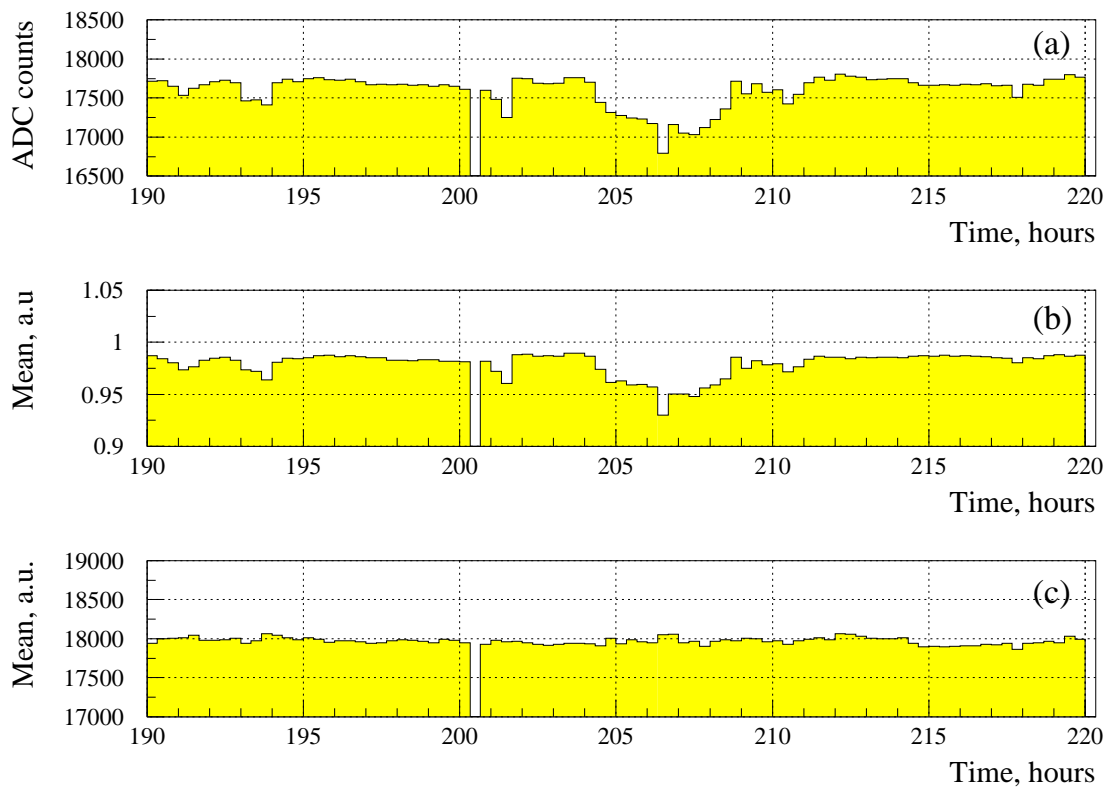}
\caption{Long-term stability histograms. Each entry is a mean value
of amplitude distribuion of induced PMT signals collected over 20 minutes
(pedestals are subtracted) from: (a) yellow led; (b) YAP light pulser;
(c) yellow LED after correction on YAP light pulser. The hole in
all three histograms corresponds to a time of no data taking.}
\label{fig:hv_cor}
\end{figure}
    
The stability of the monitoring system was estimated using the data from
the PIN photodiodes (PIN1 and PIN2) and the reference PMT ($\alpha$-PMT).
We calculated the mean pulse heights of signals from the four LED's 
measured by PIN1, PIN2 and $\alpha$-PMT, accumulated over 20-minutes intervals.

\begin{figure}
\centering
\includegraphics[width=0.95\textwidth]{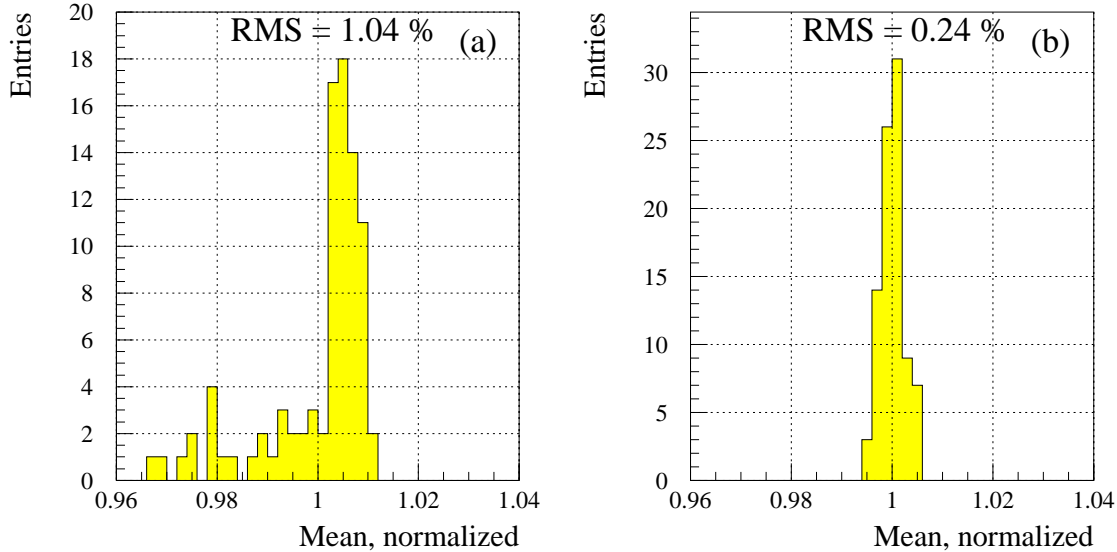}
\caption{Normalized distributions of mean amplitudes calculated for 30 hours:
(a) yellow LED in $\alpha$-PMT; the r.m.s. instability is $1.04\%$; (b) yellow LED
in $\alpha$-PMT corrected by $\alpha$-YAP light pulser; the resulting r.m.s. instability is $0.24\%$.}
\label{fig:hv_norm}
\end{figure}

Figures~\ref{fig:hv_cor}(a) and \ref{fig:hv_cor}(b) show time variation of
$\alpha$-PMT signals from the yellow LED and the $\alpha$-YAP light source over 
a 30 hour time period. The correlated variations indicate that the gain of 
the reference PMT changed during this time.
By taking the ratio between \ref{fig:hv_cor}(a) and \ref{fig:hv_cor}(b), 
shown in \ref{fig:hv_cor}(c), we can correct 
the LED intensity measured with the reference PMT for the PMT gain variation. 
Figure \ref{fig:hv_cor}(c), in fact, shows much smaller variations.
In order to evaluate the size of the variations, we formed normalized histograms of pulse 
height measurements shown in Figures \ref{fig:hv_cor}(a) and \ref{fig:hv_cor}(c), 
which are shown in Figure~\ref{fig:hv_norm}.
The r.m.s. of these distributions, expressed in percent, are 1.04\% before the PMT gain 
corrections and 0.24\% afterward.
This correction works very well, therefore all the results on LED's intensity 
variations measured by $\alpha$-PMT presented in this paper are corrected using 
the $\alpha$-YAP light source.
The r.m.s. of the corrected distribution has contributions from the variations of the LED 
intensity, as well as the statistical error 
of each measurement.
Measurements made with the PIN photodiodes reflect the LED intensity variation as well
as the variation of the PIN photodiode monitoring system.  

\section{Monitoring System Performance}

We have estimated the stability of the monitoring system for three 
continuous time intervals of different duration:\\
 a) 25 hours (short-term stability),\\
 b) 200 hours (middle-term stability),\\
 c) 2000 hours (long-term stability).\\
The results are presented in Table \ref{tab:allstab}.

\begin{table}
\begin{center}
\caption{Instability of the LED monitoring system (r.m.s.) expressed in \% over 25, 200 
and 2000 hours for the four LED's of different colors. (The first lines for the 25 h 
and 200 h intervals are the results obtained from data accumulated while the lead 
tungstate crystals were irradiated. The second ones were obtained from the recovery data.) 
The red LED data from PIN2 are missing because the pulse heights were outside the ADC range.}
\begin{tabular}{|r||c|c|c|c||c|c|c|c||c|c|c|}
\hline
&\multicolumn{4}{c||}{$\alpha$-PMT} & \multicolumn{4}{c||}{PIN1}
&\multicolumn{3}{c|}{PIN2} \\
\cline{2-12}
       & violet & blue & yellow & red & violet & blue & yellow & red
       & violet & blue & yellow \\ 
\hline\hline
25 h   &  0.38  & 0.23 &  0.28  & 0.12 & 0.37  & 0.29 &  0.16  & 0.05
       &  0.21  & 0.14 &  0.16  \\
\cline{2-12}
       &  0.29  & 0.22 &  0.19  & 0.09 & 0.33  & 0.35 &  0.15  & 0.04
       &  0.21  & 0.10 & 0.15   \\
\hline
200 h  &  0.57  & 0.46 &  0.50  & 0.36 & 0.57  & 0.50 &  0.32  & 0.08
       &  0.38  & 0.45 &  0.23  \\
\cline{2-12}
       &  0.41  & 0.33 &  0.95  & 0.72 & 0.47  & 0.38 &  0.55  & 0.07
       &  0.38  & 0.40 &  0.42  \\
\hline
2000 h &  0.54  & 0.34 &  0.95  & 0.73 & 0.79  & 0.41 &  0.55  & 0.08*
       &  0.47  & 0.50 &  0.46  \\
\hline
\end{tabular}
\label{tab:allstab}
\end{center}
* The result obtained for the first 550 hours of the recovery process.
\end{table}

\begin{figure}
\centering
\includegraphics[width=0.95\textwidth]{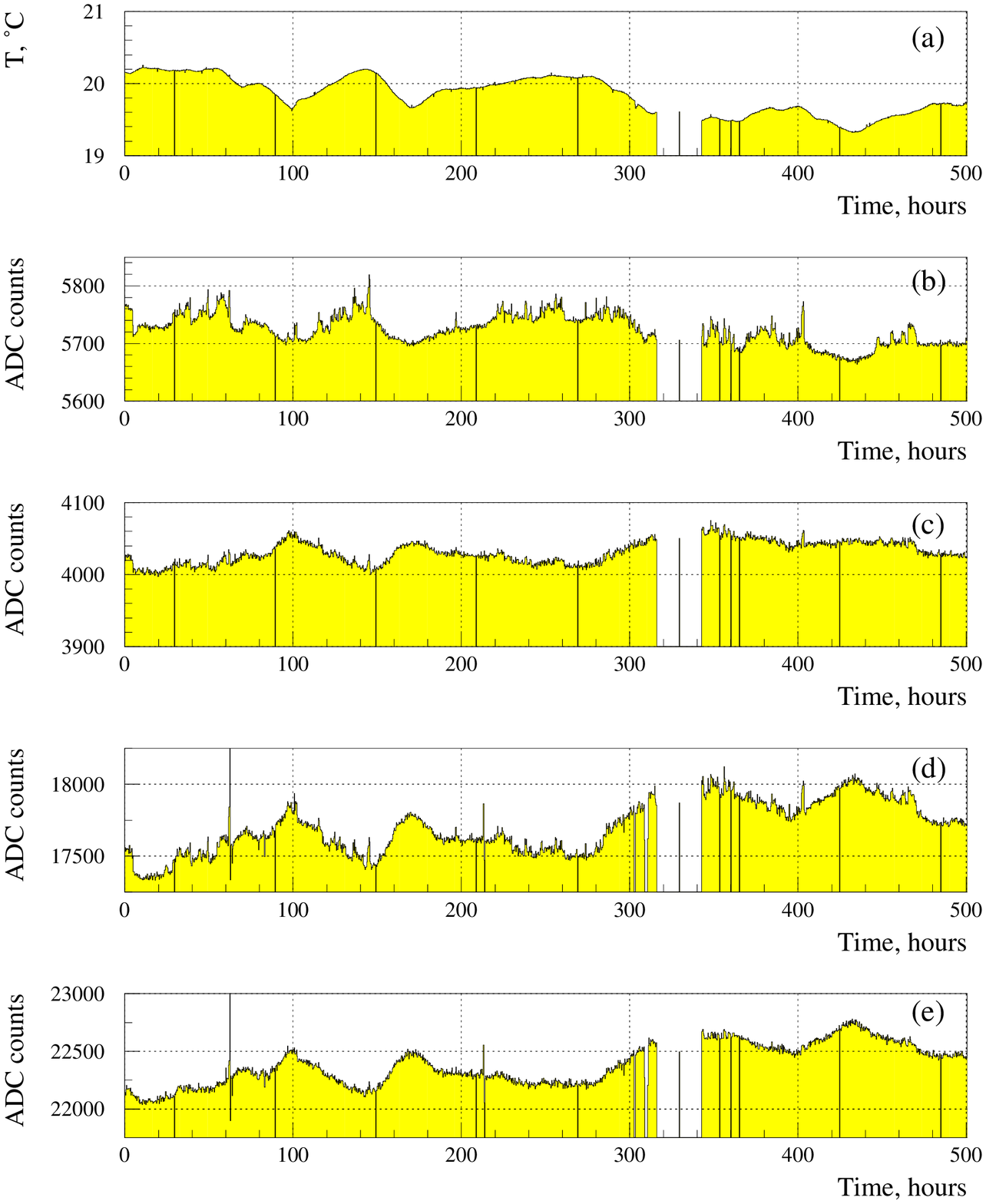}
\caption{Long-term stability histograms for the first 500 h of recovery data:
(a) temperature measured inside the box;
(b) violet LED in PIN2; (c) blue LED in PIN2; 
(d) yellow LED in $\alpha$-PMT; (e) red LED in $\alpha$-PMT.
Each entry corresponds to the average pulse height using 20 minutes of data.}
\label{fig:temper}
\end{figure}
 
The Table is organized as follows. There are two groups of results
for each of the 25 h and 200 h intervals. The first line in each set
contains the results obtained while irradiating the lead tungstate crystals; 
the second line contains the results while the crystals were recovering. 
The 2000 h interval reflects further recovery; here the data were collected 
continuously only during the first part (about 550 hours)
of this interval. After that, the HV power supply and the DAQ 
systems were turned on for only about 10-12 hours a day. We excluded from
the analysis data collected over the first 4 hours after each switch-on to 
allow the system to reach stability.
 
The results using the $\alpha$-PMT for the two different 25 h sets of data 
look very similar, but the recovery-period data appears slightly more stable. 
This may be explained by the fact that the system was somewhat affected by 
electrical noise, which was higher during accelerator operation.
The same differences are observed in the 200 h results for the violet and 
blue LED's in all three monitoring photodetectors. However, for the yellow 
and red LED's (except red in PIN1) the trend is opposite. 
It can be explained by temperature dependence which contributes the major
effect as we'll show later. 
The variations over 2000 h do not differ much from those obtained for 200 h recovery 
data. Only the violet LED results are worse over 2000 h than 200 h in all monitoring
photodetectors. This is particularly pronounced in the PIN1 data.

The sensitivity of PIN photodiodes is much better in the red region than in the blue one.
Unfortunately signals from the red LED in PIN2 were out of ADC range. 
To monitor the red LED by PIN1 we chose a fiber with bad light transmission,
thus the pulse heights in PIN1 from all other LED's are significantly smaller than those 
in PIN2. As a result, the relative statistical errors of mean pulse height calculations 
and resulting r.m.s. was higher in PIN1.

Detailed study showed that temperature variations affect the monitoring 
systems stability.
Figure~\ref{fig:temper}(a) shows the temperature variation over first 500 hours 
of the recovery process as measured by one of the sensors installed in a rather hot 
place in back of the crystal array near the PMT's. The room temperature changed 
significantly during this time interval and caused non-negligible temperature 
variation inside the box: the difference between maximum and minimum is 0.9$^\circ$C.  
Comparing the variation with time of the mean LED signals (some of them are shown in
Figure~\ref{fig:temper}) and temperature,  
we find that the violet LED signals in both PIN's are directly proportional to the
temperature, while the blue LED signal in PIN2, the red LED signal in the $\alpha$-PMT
and the yellow LED signals in all monitoring photodetectors are inversely proportional to
the temperature.
No evident dependencies on temperature were found for the violet and blue 
signals in the $\alpha$-PMT, or the blue and red signals in PIN1. 
        
\begin{figure}
\centering
\includegraphics[width=0.95\textwidth]{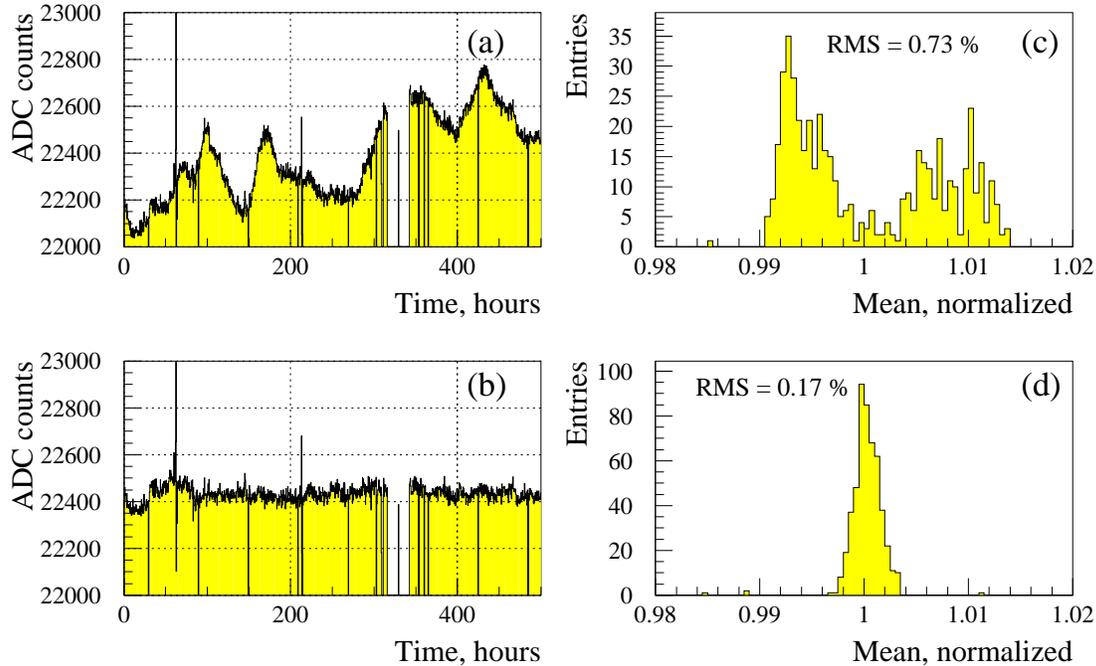}
\caption{Temperature correction of the red LED mean amplitude distribution
from  $\alpha$-PMT: (a) and (c) before correction; (b) and (d) after 
correction. The distributions (c) and (d) are obtained over 2000 hours,
while histograms (a) and (b) show the first 500 hours of this interval.}
\label{fig:tempcor}
\end{figure}

Since we didn't have temperature sensors installed near the LED pulser, the reference 
PMT and PIN's, dependence of the particular photodetector output signal on its own 
temperature cannot be obtained. 
Nevertheless, it is possible to correct for the effect of the temperature variations, since  
all the sensors inside the box gave similar curves of the temperature behavior with time.
For this purpose we plotted dependencies of the output signals in the referenced
photodetectors  on temperature measured by one of the thermo-sensors, fitted them by straight
lines and found the linear fit coefficients. After that we plotted the long-term stability
histograms with correction for the temperature variation using obtained coefficients.
Figure~\ref{fig:tempcor} illustrates the results of
applying such correction for the red LED mean amplitude distribution measured by the $\alpha$-PMT. 
Histograms \ref{fig:tempcor}(a) and \ref{fig:tempcor}(b) represent the first 500 hours of the 
recovery process, while \ref{fig:tempcor}(c) and \ref{fig:tempcor}(d) are the distributions
obtained over 2000 hours. It is clear that the temperature is
the major factor in the instability of the red LED signal. 
After applying the correction for the 
temperature variation the r.m.s. has been improved by the factor of four. 
It is interesting that signal from the same LED measured by PIN1 didn't depend 
on temperature. There are several possible explanations of this effect:\\
1) saturation of electronics;\\
2) the red LED and PIN1 don't depend on temperature;\\
3) the red LED and PIN1 in the red region have opposite temperature coefficients 
and compensate each other.\\
The linearity of electronics was checked and confirmed.
It is known from the Kingbright LED data sheets \cite{web}, that light 
intensities of the violet, yellow and red LED's have negative temperature coefficients.
Therefore the last hypothesis seems to be the most reasonable. Also it might explain, 
why the violet LED signals in both PIN's behave in opposite manner compared to other 
color LED's: temperature coefficient of the PIN in violet region is greater than that 
of the violet LED and has opposite sign. 

The r.m.s. values of mean signal distributions, corrected for temperature variations, 
are presented in Table \ref{tab:stabcor}.
The results are given only for those combinations LED -- photodetector that showed
temperature dependencies.  
 
\begin{table}
\begin{center}
\caption{Instability of the LED monitoring system (r.m.s.) expressed in \% measured using 
data intervals of either
200 or 2000 hours and corrected for temperature dependence. (The first line for
200 h interval represents results obtained from
data accumulated during lead tungstate crystal irradiation and  
the second one from the recovery data.) }
\begin{tabular}{|r||c|c|c|c||c|c|c|c||c|c|c|}
\hline
&\multicolumn{4}{c||}{$\alpha$-PMT} & \multicolumn{4}{c||}{PIN1}
&\multicolumn{3}{c|}{PIN2} \\
\cline{2-12}
       & violet & blue & yellow & red & violet & blue & yellow & red
       & violet & blue & yellow \\ 
\hline\hline
200 h  &    -   &  -   &  0.37  & 0.21 & 0.55  &  -   &  0.27  & -
       &  0.28  & 0.42 &  0.22  \\
\cline{2-12}
       &    -   &  -   &  0.29  & 0.16 & 0.43  &  -   &  0.23  & -
       &  0.24  & 0.13 &  0.19  \\
\hline
2000 h &    -   &  -   &  0.29  & 0.17 & 0.72  &  -   &  0.29  &  -
       &  0.37  & 0.34 &  0.26  \\
\hline
\end{tabular}
\label{tab:stabcor}
\end{center}
\end{table} 

\section{Summary}

A light monitoring system with four LEDs of different wavelengths
has been designed for the BTeV lead tungstate electromagnetic 
calorimeter prototype and assembled and tested in a test-beam at Protvino.
The system provided an individual check of each prototype channel
by monitoring the PMT's gain variation and crystals transparency change due 
to the beam irradiation.
  
Each color LED was fed to the crystal-PMT combination using a light fiber.
In addition, the LED's were connected to 
two PIN photodiodes and a PMT with calibrated light source placed on its 
window, to provide reference signals.
We have analyzed the stability of the signals produced by each combination LED --
monitoring photodetector for several time intervals of different duration.
This analysis allowed us to determine, which  photodetector gives
the most stable reference signal for any particular LED. 
The best combinations were stable within:
\begin{itemize}
\item $0.2\%$ over one day;
\item $0.4-0.7\%$ over one week and longer (up to 3 months).
\end{itemize}
Variations of temperature were found to be the most important factor which affected the
monitoring system performance. The correction for temperature was included 
in the off-line analysis in order to estimate the long-term 
stability of the system in condition of stable temperature. This correction
decreased the r.m.s. instability to $0.2-0.4 \%$ over several months.

The system completely satisfies our demands on stability for the methodical
tests of crystals radiation hardness. It allows us to perform measurements 
with an accuracy of $0.7\%$ over a few months. Moreover, the
24-hour system performance has already exceeded the requirements for BTeV.
We are going to design an LED monitoring system for
the BTeV EMCAL with the required long-term stability of $0.2\%$ over a week using
the same technical solution in the part of LED driver. Special care will be put 
in the choice of LED's and monitoring photodiodes as well as the good 
temperature stabilization of the system.

\section{Acknowledgments}
    We thank the IHEP management for providing us infrastructure support
and accelerator time. 
We thank Sten Hansen for developing and
producing the PMT amplifiers for our studies.
Special thanks to Fermilab for providing equipment for data acquisition.
This work was partially supported by the U.S.
National Science Foundation and the Department of Energy
as well as the Russian Foundation for Basic Research grant 02-02-39008.

\end{document}